\input epsf.tex
\documentstyle[twocolumn,prb,aps]{revtex}
\begin{document}
\twocolumn[
\title{Semiclassical description of spin ladders}
\author{D. S\'en\'echal}
\address{Centre de Recherche en Physique du Solide et
D\'epartement de Physique,}
\address{Universit\'e de Sherbrooke, Sherbrooke, Qu\'ebec, Canada J1K 2R1.}
\date{June 1995}
\maketitle

\widetext\leftskip=1cm\rightskip=1cm\nointerlineskip\small
The Heisenberg spin ladder is studied in the semiclassical limit, via
a mapping to the nonlinear $\sigma$ model. Different treatments are
needed if the inter-chain coupling $K$ is small, intermediate or large.
For intermediate coupling a single nonlinear $\sigma$ model is used for the
ladder. Its predicts a spin gap for all nonzero values of $K$ if the sum
$s+\tilde s$ of the spins of the two chains is an integer, and no gap
otherwise. For small $K$, a better treatment proceeds by coupling two
nonlinear sigma models, one for each chain. For integer $s=\tilde s$, the
saddle-point approximation predicts a sharp drop in the gap as $K$ increases
from zero. A Monte-Carlo simulation of a spin 1 ladder is presented which
supports the analytical results.
]
\narrowtext
\section{Introduction}
Interest in one-dimensional quantum antiferromagnets has been great ever
since Haldane conjectured\cite{Haldane83} that integer-spin chains have a gap
in their excitation spectrum while half-integer spin chains do not. More
recently, coupled spin chains have aroused some interest, in particular the
so-called Heisenberg ladder, whose Hamiltonian may be written as follows:
\begin{equation}
\label{Hamil1}
H = J\sum_{i}\left\{ {\bf S}_i\cdot{\bf S}_{i+1} +
{\bf\tilde S}_i\cdot{\bf\tilde S}_{i+1}\right\} +
K\sum_{i}{\bf S}_i\cdot{\bf\tilde S}_i
\end{equation}
wherein ${\bf S}_i$ and ${\bf\tilde S}_i$ are the spin operators at site
$i$ for chains A and B respectively. The relevant parameters are the
spins $s$ and $\tilde s$ of each chain, and the coupling ratio
$K/J$. Since this system is one-dimensional and is characterized by a
continuous order parameter, one does not expect any long range order for
any value of the couplings $J$ and $K$. The important question is rather
how the spin excitation gap $\Delta$ develops or varies as the inter-chain
coupling $K$ is turned on. The case of two spin-$1\over2$ chains is the one
that has received attention recently, and has been studied with various
techniques:  Bosonization\cite{Strong92}, exact
diagonalization\cite{Dagotto92,Barnes93}, Quantum Monte-Carlo\cite{Barnes93}
and density-matrix renormalization\cite{Azzouz94,White95}. The prevailing
conclusion is that a gap $\Delta$ appears for any nonzero inter-chain coupling
$K$. This gap increases with $K$ in an almost linear fashion.

In this work we will study the Heisenberg ladder with
semi-classical techniques, i.e., by using an approximate mapping from the
Hamiltonian to the nonlinear $\sigma$ model. As long as $K$ and $J$ are
not too different, one may assume short range antiferromagnetic (AF) order
along and accross the chains, and work out a direct mapping with the
nonlinear $\sigma$ model. This will be described in section II. However,
the validity of such a mapping is questionable when $K/J$ is too small (or too
large). If $K\ll J$ one should rather consider two coupled nonlinear $\sigma$
models, one for each chain. For spin-$1\over2$ chains this analysis is
complicated by the existence of topological terms, and we shall carry it only
for integer spin; this is done in section III. In section IV we will compare
these semi-classical results with the outcome of a quantum Monte-Carlo
simulation of the spin-1 Heisenberg ladder.

\section{Intermediate coupling analysis}

\subsection{Derivation of the $\sigma$ model}

In this section we will show how the AF Heisenberg ladder defined by
the Hamiltonian (\ref{Hamil1}) may be described, in the continuum
and semi-classical limit, by the one-dimensional nonlinear $\sigma$ model
with Lagrangian density
\begin{equation}
\label{NLs}
{\cal L}_\sigma = {1\over 2g}\left\{
{1\over v} (\partial_t{\bf m})^2 - v (\partial_x{\bf m})^2 \right\}
\end{equation}
wherein the field ${\bf m}(x,t)$ is a unit vector.
The coupling $g$ and the velocity $v$ obtained semiclassically depend
on the spins $s$ and $\tilde s$, and on the inter-chain and intra-chain
couplings $K$ and $J$. If the two chains are alike
($s=\tilde s$), these parameters are
\begin{equation}
\label{vg}
v = 2asJ\sqrt{1+K/2J}\qquad g = {1\over s}\sqrt{1+K/2J}
\end{equation}
If $s\ne\tilde s$, the corresponding expressions are more complicated.

In addition to the $\sigma$ model Lagrangian (\ref{NLs}), there
also appears a topological term
\begin{equation}
{\cal L}_{top} = {\theta\over4\pi} {\bf m}\cdot(\partial_t{\bf m}\times
\partial_x{\bf m})
\end{equation}
with $\theta=2\pi(\tilde s-s)$. One easily shows that, with suitable
boundary conditions at infinity, the action obtained by integrating
${\cal L}_{top}$ is an integer times $\theta$. It follows that this
topological term has no effect if $\tilde s-s$ is an integer. On the
other hand, if $\tilde s-s$ is a half-integer, i.e., if one of the chains
is made of half-integer spins and the other of integer spins, then the
topological term appears as it does in the semiclassical analysis of
a single half-integer spin chain, from which we conjecture, following
Haldane, that such a system does not have a gap.

Let us now go through this semiclassical derivation. We will proceed as
in Ref.~\onlinecite{Senechal93B}. The idea is to first describe the dynamics
of a single spin in the Lagrangian formalism, with the help of spin coherent
states. For details on this technique the reader is referred to Fradkin's
text\cite{Fradkin91} or to the review by Manousakis\cite{Manousakis91}.
The dynamics of a single spin $\bf S$ without interaction is described
by a unit vector $\bf n$ such that ${\bf S} = s{\bf n}$ and the
corresponding action, named after Wess and Zumino (WZ), is nonlocal:
\begin{equation}
\label{actionWZ}
S_{WZ} = \int_0^T dt~{\cal L}_{kin} =
s\int_0^1 d\tau\int_0^T dt~ {\bf n}\cdot(\partial_\tau{\bf n}
\times \partial_t{\bf n})
\end{equation}
Here the time $t$ runs from 0 to some finite period $T$ and $\tau$ is an
auxiliary coordinate introduced in order to parametrize, along with $t$, the
spherical cap delimited by the curve  ${\bf n}(t)$ from $t=0$ to $t=T$.
What is actually needed is the variation of this action upon a small
change $\delta{\bf n}$:
\begin{equation}
\label{variaWZ}
\delta S_{WZ} = s\int_0^T dt~ \delta{\bf n}\cdot({\bf n}
\times \partial_t{\bf n})
\end{equation}

Each spin on the ladder may be described by a unit vector $\bf n$ like
above.  We expect short range AF order, with a `magnetic cell'
containing four spins. If we call ${\bf r}$ the position of a reference spin
in each magnetic cell, the positions ${\bf r}'$ of each of the four spins
within the cell may be written as
\begin{equation}
{\bf r}' = {\bf r} + \alpha{\bf\hat x}+\beta{\bf\hat y}
\end{equation}
where ${\bf\hat x}$ and ${\bf\hat y}$ are the lattice vectors
(respectively along and across the ladder) and where
$\alpha$ and $\beta$ each run from 0 to 1.
The short range AF order motivates the introduction of a unit vector
$\bf n$ which varies slowly in space:
\begin{eqnarray}
\label{unitvec}
&&{\bf S}({\bf r}) = s{\bf n}({\bf r}) \nonumber\\
&&{\bf S}({\bf r}+{\bf\hat x}) = -s{\bf n}({\bf r}+{\bf\hat x}) \nonumber\\
&&{\bf\tilde S}({\bf r}+{\bf\hat y}) =
-\tilde s{\bf n}({\bf r}+{\bf\hat y})\nonumber\\
&&{\bf\tilde S}({\bf r}+{\bf\hat x}+{\bf\hat y}) =
\tilde s{\bf n}({\bf r}+{\bf\hat x}+{\bf\hat y})
\end{eqnarray}
The four spins of the unit cell may be described differently, at our
convenience, by four different vectors, which we choose as follows (in
terms of the unit vector defined above):
\begin{eqnarray}
\label{auxFields}
{\bf n}({\bf r}) &=&
{\bf m}({\bf r}) + a\left[\phantom{-}{\bf l}_{01}({\bf r})+
{\bf l}_{10}({\bf r})+{\bf l}_{11}({\bf r})\right]\nonumber\\
{\bf n}({\bf r}+{\bf\hat x}) &=&
{\bf m}({\bf r}) + a\left[\phantom{-}{\bf l}_{01}({\bf r})-
{\bf l}_{10}({\bf r})-{\bf l}_{11}({\bf r})\right]\nonumber\\
{\bf n}({\bf r}+{\bf\hat y}) &=&
{\bf m}({\bf r}) + a\left[-{\bf l}_{01}({\bf r})+
{\bf l}_{10}({\bf r})-{\bf l}_{11}({\bf r})\right]\nonumber\\
{\bf n}({\bf r}+{\bf\hat x}+{\bf\hat y}) &=&
{\bf m}({\bf r}) + a\left[-{\bf l}_{01}({\bf r})-
{\bf l}_{10}({\bf r})+{\bf l}_{11}({\bf r})\right]
\end{eqnarray}
The mean value of ${\bf n}$ within a magnetic cell is ${\bf m}$. In
addition, we have introduced three deviation field ${\bf l}_{10}$,
${\bf l}_{01}$ and ${\bf l}_{11}$
which describe deviations from strict AF order. We have
included a factor of the lattice spacing $a$ in front of these deviation
fields in order to emphasize our assumption that the deviations from the
short-range AF order are small, and to control this approximation in the same
way as the continuum limit. The four fields ${\bf m}$ and ${\bf l}_{ij}$
all depend on the cell position ${\bf r} = x{\bf\hat x}$ only, since they
are defined for a cell as a whole, and are entirely equivalent to the
specification of the four spins ${\bf S}({\bf r}')$ of that cell. They
are not independent, but must obey the constraints imposed by the relation
${\bf n}^2=1$. Thus, we have $4\times2=8$ degrees of freedom per unit cell.

So far the treatment has been exact. Now we will express the kinetic term
$S_{kin}$ (the sum of the Wess-Zumino actions for each of the spins) and
the potential term (the Heisenberg Hamiltonian) in terms of these new
variables, to lowest nontrivial order in $a$. We will then integrate out (in
the functional sense) the deviation fields to end up with a theory defined
only in terms of the staggered magnetization $\bf m$.

Let us start with the kinetic term. In principle this term is nonlocal,
as expressed in Eq.~(\ref{actionWZ}). The local AF order allows us to
write it in local form (to first order in $a$) since the WZ action is
odd under the reversal of spin and Eq.~(\ref{variaWZ}) may be applied
between two neighboring sites. If we define the difference
\begin{equation}
\delta_x{\bf n}({\bf r}') =
{\bf n}({\bf r}'+{\bf\hat x})-{\bf n}({\bf r}')
\end{equation}
(likewise for $\delta_y{\bf n}$)
one may write the kinetic term, to first order in $a$, as
\begin{equation}
S_{kin} = \sum_{\bf r}\left[
-s\delta_x{\bf n}({\bf r})
+\tilde s\delta_x{\bf n}({\bf r}+{\bf\hat y})\right]
\cdot({\bf m}\times\partial_t {\bf m})
\end{equation}
Since $\delta_x{\bf n}({\bf r})=-2a(l_{10}+l_{11})$ and
$\delta_x{\bf n}({\bf r}+{\bf\hat y})=-2a(l_{10}-l_{11})$, this reduces,
after replacing the sum by an integral, to the following:
\begin{equation}
S_{kin} = \int dx dt~
\left[(s+\tilde s){\bf l}_{11}+(s-\tilde s){\bf l}_{10}\right]
\cdot({\bf m}\times\partial_t{\bf m})
\end{equation}

The Heisenberg Hamiltonian may be separated into three parts:
$H=V_x+\tilde V_x + V_y$ where, up to an additive constant,
\begin{mathletters}
\begin{eqnarray}
\label{pot1}
 V_x &=& {1\over2} s^2 J \sum_{\bf r} \Big\{
[\delta_x {\bf n}({\bf r})]^2+[\delta_x {\bf n}({\bf r}+{\bf\hat x})]^2
\Big\} \\
 \tilde V_x &=& {1\over2} \tilde s^2 J \sum_{\bf r} \Big\{
[\delta_x {\bf n}({\bf r}+{\bf\hat y})]^2+
[\delta_x {\bf n}({\bf r}+{\bf\hat y}+{\bf\hat x})]^2
\Big\} \\
 V_y &=& {1\over2} \tilde ss K \sum_{\bf r} \Big\{
[\delta_y {\bf n}({\bf r})]^2+[\delta_y {\bf n}({\bf r}+{\bf\hat x})]^2
\Big\}
\end{eqnarray}
\end{mathletters}
Expressing the above in terms of the fields defined in (\ref{auxFields}),
we perform a first order Taylor expansion on ${\bf m}$, while we
essentially neglect the spatial variations of the deviation fields. For
instance, we have
\begin{equation}
\delta_x {\bf n}({\bf r}+{\bf\hat y}+{\bf\hat x}) =
2a\partial_x{\bf m} + 2a({\bf l}_{10}-{\bf l}_{11})
\end{equation}
Converting the sums of Eq.~(\ref{pot1}) into integrals, we find the
following Hamiltonian density
\begin{eqnarray}
\label{Hamil2}
{\cal H}&=& 2Jas^2\left[ {1\over2}(\partial_x {\bf m})^2 +
({\bf l}_{10}+{\bf l}_{11})\cdot({\bf l}_{10}+{\bf l}_{11}+
\partial_x {\bf m})\right] \nonumber\\
&+& 2Ja\tilde s^2\left[ {1\over2}(\partial_x {\bf m})^2 +
({\bf l}_{10}-{\bf l}_{11})\cdot({\bf l}_{10}-{\bf l}_{11}+
\partial_x {\bf m})\right] \nonumber\\
&+& 2Ka\tilde ss({\bf l}_{01}^2+{\bf l}_{11}^2)
\end{eqnarray}
The complete Lagrangian density is of course
${\cal L} = {\cal L}_{kin} - {\cal H}$.

The next step is the functional integration of the deviation fields.
Since these fields occur at most in a quadratic fashion, they may
be integrated simply by solving the classical equations of motion
and substituting the solutions back into the Lagrangian density. Taking
variations of $\cal L$ with respect to ${\bf l}_{10}$, ${\bf l}_{01}$
and ${\bf l}_{11}$, one finds, after some algebra,
\begin{mathletters}
\label{classEqs}
\begin{eqnarray}
l_{11} &=& {1\over 4aJ}{s+\tilde s\over s\tilde s}
({\bf m}\times\partial_t{\bf m})\\
l_{01} &=& 0 \\
l_{10} &=& -{1\over2}\partial_x{\bf m}+ ({\bf m}\times\partial_t{\bf m})
{\tilde s-s\over 4Ja s\tilde s}\left\{
1 + {K\over 2J}{(s+\tilde s)^2\over s\tilde s}\right\}
\end{eqnarray}
\end{mathletters}
Substituting into the expression for $\cal L$, one finds
\begin{equation}
{\cal L} = {1\over2}(\tilde s-s)
{\bf m}\cdot(\partial_t{\bf m}\times\partial_x{\bf m})
+ A(\partial_t{\bf m})^2 - B(\partial_x{\bf m})^2
\end{equation}
where the constants $A$ and $B$ are
\begin{mathletters}
\begin{eqnarray}
A &=& {1\over 4aJ}{1+(K/2J)(s-\tilde s)^2/4s\tilde s \over
1+(K/2J)(s^2+\tilde s^2)/2s\tilde s} \\
B &=& \frac12 aJ(s^2+\tilde s^2)
\end{eqnarray}
\end{mathletters}

Of course, a simple substitution of Eqs (\ref{classEqs}) into $\cal L$ may
seem dishonest since, as we indicated earlier, the deviations fields are
not independent, but constrained by the fixed length of each spin.
Expressed in terms of $\bf m$ and ${\bf l}_{ij}$, these
constraints are
\begin{mathletters}
\label{constraints}
\begin{eqnarray}
 {\bf m}^2 + a^2({\bf l}_{01}^2+{\bf l}_{10}^2+{\bf l}_{11}^2) &=& 1 \\
 {\bf m}\cdot{\bf l}_{01} + a {\bf l}_{10}\cdot {\bf l}_{11} &=& 0 \\
 {\bf m}\cdot{\bf l}_{10} + a {\bf l}_{01}\cdot {\bf l}_{11} &=& 0 \\
 {\bf m}\cdot{\bf l}_{11} + a {\bf l}_{10}\cdot {\bf l}_{01} &=& 0
\end{eqnarray}
\end{mathletters}
Since the classical equations (\ref{classEqs}) are valid only at lowest
order in $a$, it suffices to apply the above constraint equations at that
order:
\begin{equation}
{\bf m}^2 = 1 \qquad\qquad {\bf m}\cdot{\bf l}_{ij} = 0
\end{equation}
In this form the constraints are entirely compatible with the equations
(\ref{classEqs}) and we may forget our misgivings. The constraint
${\bf m}^2 = 1$ is part of the definition of the model (\ref{NLs}).
The characteristic velocity of Eq.~(\ref{NLs}) is then
$v=\sqrt{B/A}$ and the coupling is $g=1/2Av$. The coefficient
of the topological term is indeed $\theta=2\pi(\tilde s-s)$, as
announced above. It is easily checked that the correct expressions for $v$
and $g$ in the case $s=\tilde s$ coincide with Eq.~(\ref{vg}).
This concludes the semi-classical derivation of the nonlinear model for the
Heisenberg ladder.

\subsection{Estimating the gap}
The $O(3)/O(2)$ nonlinear $\sigma$ model -- as the model defined in
(\ref{NLs}) is precisely known -- lends itself to an estimate of the
spin excitation gap in the saddle-point approximation. This has been
described in the literature (cf. the reviews in Refs.
\onlinecite{Affleck89} and \onlinecite{Manousakis91}) but will be
summarized briefly here.

If it were not for the constraint ${\bf m}^2=1$, the model defined
by the Lagrangian (\ref{NLs}) would simply describe free, massless
excitations. A classical analysis of the model (incorporating the
constraint) reveals a spontaneous breaking of the internal rotation symmetry
to a ground state in which the field $\bf m$ is uniform. However, this
ordered state is destroyed by quantum fluctuations in dimension 1.
This is revealed by the following saddle-point analysis: First, one
implements the unit-vector constraint by introducing a Lagrange multiplier
$\sigma(x,t)$ at every space-time point and by adding the term
$\sigma({\bf m}^2-1)$ to the Lagrangian density. After rescaling the fields
and going to imaginary time $\tau=-it$, the complete Lagrangian
may be written as
\begin{equation}
\label{Lag2}
{\cal L}_E = \frac12 (\partial{\bf m})^2
+ \frac12\sigma({\bf m}^2-1/g)
\end{equation}
Here $(\partial{\bf m})^2$ stands for
$(\partial_\tau{\bf m})^2 + (\partial_x{\bf m})^2$.
We have set the velocity $v$ equal to one and will restore it later by
dimensional analysis. The field $\bf m$ is now unconstrained, and its
functional integration yields the effective potential $V(\sigma)$ for the
multiplier field $\sigma$. In the saddle-point approximation, the
fluctuations of $\sigma$ are neglected ($\sigma$ is then a constant) and the
derivative of the effective potential is
\begin{eqnarray}
V'(\sigma) &=& \frac1{2g} - \frac32\int {dk\over2\pi}{d\omega\over2\pi}
{1\over \omega^2+k^2+\sigma} \nonumber\\
&=& \frac1{2g} + \frac3{8\pi}\ln(\sigma/\Lambda^2)
\end{eqnarray}
wherein $\Lambda$ is a momentum cutoff, such that $\Lambda a\sim1$.
The saddle-point is thus fixed by the vanishing of the above, which
occurs at $\sigma_0=\Lambda^2\exp-(4\pi/3g)$. A nonzero value of the
saddle-point means that the field $\sigma$ should be redefined with
respect to the position of the saddle, as $\sigma=\tilde\sigma+\sigma_0$,
wherein $\tilde\sigma$ is a field fluctuating about zero and $\sigma_0$
is a constant multiplying $\frac12{\bf m}^2$ in the quantum effective action
of the nonlinear $\sigma$ model. Thus, $\sigma_0$ has the interpretation
of a quantum fluctuation-induced mass squared for the field ${\bf m}$.
At this level of approximation we consider no other correction to the
effective action: in the language of the $1/N$ expansion, we stay at
lowest order. Restoring the characterstic velocity, the excitation gap
may be expressed as
\begin{equation}
\label{gap1}
\Delta = \Lambda v\exp-(2\pi/3g)~.
\end{equation}
Note that this expression is `nonperturbative', in the sense that it admits
no series expansion in powers of $g$ about $g=0$. Since $g\sim 1/s$ is the
natural expansion parameter of spin-wave theory, it is quite understandable
that the latter cannot account for the existence of the Haldane gap.

\begin{figure}[tpb]
\vglue 0.4cm\epsfxsize 8cm\centerline{\epsfbox{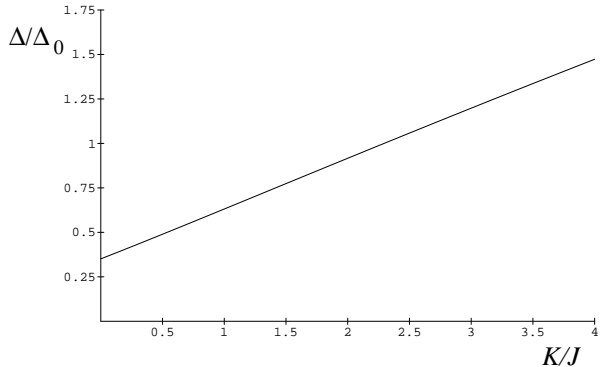}}\vglue 0.4cm
\caption{
Reduced gap $\Delta/\Delta_0$ as a function of inter-chain coupling $K/J$,
as obtained in the saddle-point approximation of the non-linear sigma
model. The value of $\Delta_0$ comes from the saddle-point approximation in
the single-chain nonlinear sigma model.}
\label{fig1}
\end{figure}
If we now substitute the expression (\ref{vg}) for $v$ and $g$, we find
\begin{equation}
\label{gap2}
\Delta/J = 2(\Lambda a)s\sqrt{1+K/2J}\exp-{2\pi s\over 3\sqrt{1+K/2J}}
\end{equation}
This formula indicates a growth of $\Delta$ as a function of $K$, due
to the variation of both $g$ and $v$. Since the model is defined on the
continuum, an arbitrary lattice spacing $a$ has been introduced and the
precise value of the gap cannot be known. However, it is hoped that its
dependence on various parameters may be well approximated by this
method. The $K$-dependence of the gap $\Delta$ obtained in this fashion is
illustrated on Fig.~\ref{fig1}. Up to fairly large values of $K$, the
dependence is almost linear. The small $K$ behavior is grossly wrong,
as will be discussed below.

\subsection{Extensions}
The calculation of part IIA may also be applied to the case of a
ferromagnetic inter-chain coupling ($K<0$). We simply have to
redefine the spins of the second chain by a sign:
${\bf\tilde S}_i\to-{\bf\tilde S}_i$. Of course Eq.~(\ref{unitvec}) has
to be modified (the last two r.h.s. change sign) but Eq.~(\ref{auxFields})
stays the same. We find that the Hamiltonian density (\ref{Hamil2}) is
unchanged, but the kinetic term is modified to
\begin{equation}
{\cal L}_{kin} = 2s{\bf l}_{10}\cdot({\bf m}\times\partial_t{\bf m})
\end{equation}
(we set $s=\tilde s$ for simplicity).
Integrating out the deviation fields, we find in this case that all
traces of $K$ disappear! We again find the nonlinear $\sigma$ model
Lagrangian (\ref{NLs}), but this time with the parameters
\begin{equation}
v = 2Jas \qquad \quad g = {1\over s}\qquad(K<0)
\end{equation}
If we compare with the corresponding parameters for a single isolated chain
($v=2Jas$ and $g=2/s$), we see that the ladder behaves as if it were a
single chain of effective spin $2s$, albeit with a characteristic velocity
half as large, which implies a gap half as large as that of the single chain.
In other words, this approach implies that the spin-$\frac12$ ladder with
ferromagnetic inter-chain coupling should behave like a single spin-1 chain,
but with half the usual Haldane gap: $\Delta\approx0.205J$, for all
values of $K$.  Of course, like everything in this section,
this result is not to be trusted when $K\ll J$. However, it has been
confirmed by density-matrix renormalization group calculations\cite{Chen94}
for reasonable values of $K$.

Another possible extension of the above analysis is a variant of the Kondo
necklace, wherein the coupling is taken to be Heisenberg-like. The
corresponding Hamiltonian is obtained from (\ref{Hamil1}) by deleting the
${\bf S}_i\cdot {\bf S}_{i+1}$ term. In this system we also expect short
range AF order and we may use the same decomposition (\ref{auxFields}).
Here the Hamiltonian density will be simpler (the second line of
Eq.~(\ref{Hamil2}) will disappear) with the end result $v=\sqrt{JK}\,as$
and $g=(1/s)\sqrt{K/J}$.

\section{Weak coupling analysis}

The results of the previous section are no longer applicable if $K\ll J$.
Indeed, some of the heuristic assumptions made in the derivation of the
nonlinear $\sigma$ model are violated: a small inter-chain coupling implies
that inter-chain AF correlations are weaker and the deviation field
$a{\bf l}_{01}$ is not necessarily small. We may guess the rough effect this
will have on the gap by the following crude argument. Assuming that the
nonlinear $\sigma$ model is still applicable, the fact that $a{\bf l}_{01}$
is not small alters the constraint on the staggered magnetization. After
rescaling the fields as in Eq.~(\ref{Lag2}), this constraint would be
${\bf m}^2 = 1/g_{eff} <1/g$. The `effective coupling' $g_{eff}$ is greater
than $g$, and accordingly the gap is larger, as implied by Eq.~(\ref{gap1}).
We thus would expect the actual gap to be greater than what is predicted by
Eq.~(\ref{gap2}), as $K$ approaches zero. In fact, the gap obtained for a
single spin chain in the saddle-point approximation of the nonlinear sigma
model is $\Delta/J = 2s(\Lambda a) e^{-\pi s/3}$, whereas the $K\to0$ limit of
Eq.~(\ref{gap2}) is smaller by a factor $e^{\pi s/3}$. Of course, this holds
for integer spin only. For half-integer spin chains, the gap should disappear
as $K\to0$, according to the Haldane conjecture.

Since the deviation $a{\bf l}_{01}$ is not necessarily small at weak coupling,
the constraints (\ref{constraints}) can no longer be simplified and the
integration of the deviation fields is no longer straightforward. Then we
might as well treat the two chains separately in the semi-classical
approximation and incorporate the inter-chain coupling afterwards. Each chain
is then described by its own staggered magnetization field ${\bf m}_i$
($i=1,2$) and its own nonlinear $\sigma$ model with Lagrangian density
(\ref{NLs}). After rescaling the fields as in Eq.~(\ref{Lag2}), we find that
the inter-chain interaction takes the form
${\cal L}_1 = h{\bf m}_1\cdot{\bf m}_2$, with $h=2Ks/a$. In order to estimate
the gap let us, as before, introduce Lagrange multiplier fields for the
constraints ${\bf m}_1^2={\bf m}_2^2=1$. The total Lagrangian density in
imaginary time now reads
\begin{equation}
\label{coupledLag}
{\cal L}_E = \frac12\sum_{i=1}^2\left[(\partial{\bf m}_i)^2
+ \sigma_i({\bf m}_i^2-1/g)\right] + h{\bf m}_1\cdot{\bf m}_2
\end{equation}
We have again set the characteristic velocity to 1. We should now integrate
out ${\bf m}_1$ and ${\bf m}_2$, except that we are dealing here with a
`mass matrix'
\begin{equation}
\label{massMatrix}
\Sigma = \pmatrix{\sigma_1 & h\cr h & \sigma_2\cr}
\end{equation}
The eigenvalues of this mass matrix are
\begin{equation}
\label{eigen}
\lambda_\pm = \frac12\left\{
\sigma_1 + \sigma_2 \pm \sqrt{(\sigma_1-\sigma_2)^2+4h^2}\right\}
\end{equation}

The effective potential $V(\sigma_1,\sigma_2)$ obtained by integrating
the fields ${\bf m}_{1,2}$ is
\begin{equation}
V(\sigma_1,\sigma_2) = {1\over g}(\sigma_1+\sigma_2)
-\frac32\int {dk\over2\pi}{d\omega\over2\pi}
{\rm tr}\ln\left[\omega^2+k^2+\Sigma\right]
\end{equation}
The saddle-point equations $\partial V/\partial\sigma_1=0$ and
$\partial V/\partial\sigma_2=0$ may be expressed as
\begin{eqnarray}
{1\over g} &=& {3\over 4\pi}\ln{\Lambda^2\over\sqrt{\lambda_+\lambda_-}} \\
0 &=& {3\over16\pi}{\sigma_1-\sigma_2\over\lambda_+-\lambda_-}
\ln{\lambda_+\over\lambda_-}
\end{eqnarray}
The solution to these equations is $\sigma_1=\sigma_2=\sigma$ and
$\lambda_\pm = \sigma\pm h$, with
\begin{equation}
\sigma = \sqrt{\sigma_0+h^2}
\end{equation}
where $\sigma_0$ is the $h=0$ value. The square of the gap to the lowest
excited state being $\lambda_-$, this gap, relative to its $h=0$ value, is
\begin{equation}
\label{gapM1}
{\Delta\over\Delta_0} = \sqrt{\sqrt{1+\bar h^2}-\bar h}
\end{equation}
where $\bar h = h/\Delta_0^2$. If we substitute for $\Delta_0$ the
saddle-point value and the known numerical value $0.41 J$, we find
$\bar h\approx 23.8 K/J$. According to the relation (\ref{gapM1}), the gap
decreases linearly as $K/J$ increases and then flattens towards zero. Because
of the large numerical factor ($\sim$ 23.8) between $\bar h$ and $K/J$, this
drop is quite rapid. Note that Eq.~(\ref{gapM1}) is obviously wrong when
$K/J$ is too large.

\begin{figure}[tpb]
\vglue 0.4cm\epsfxsize 8cm\centerline{\epsfbox{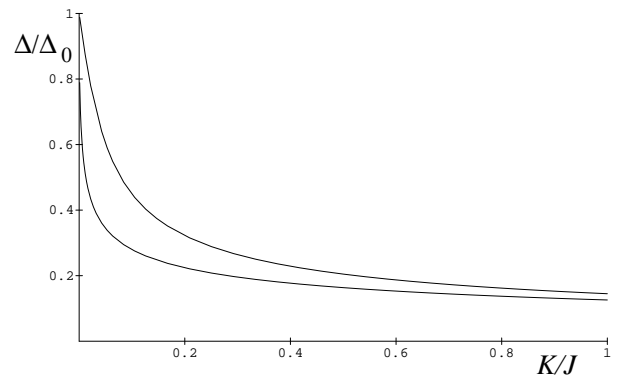}}\vglue 0.4cm
\caption{
Reduced gap $\Delta/\Delta_0$ as a function of inter-chain coupling $K/J$,
as obtained in the saddle-point approximation of the coupled non-linear
sigma models. The upper curve illustrates Eq.~(24), while the lower
curve is a numerical solution of Eq.~(41), obtained in a different
saddle-point approximation.}
\label{fig2}
\end{figure}

The saddle-point result (\ref{gapM1}) will receive corrections in
$1/N$. We shall not get into these calculations. Instead, let us
show how a slightly different result may be obtained by applying
the saddle-point approximation in terms of the masses square $\lambda_\pm$
instead of $\sigma_{1,2}$. Let us make a linear transformation within the
internal space spanned by the fields ${\bf m}_1$ and ${\bf m}_2$ such as to
make the mass matrix $\Sigma$ diagonal.
We proceed to a functional change of integration variables, replacing
$\sigma_i$ by $\lambda_\pm$, and ${\bf m}_i$ by ${\bf m}_\pm$. The Jacobian
associated with this change of variables is trivial in the latter case, but
not in the former:
\begin{equation}
{\rm det}\left\vert {\partial\sigma_i\over\partial\lambda_j}\right\vert
={\lambda_+-\lambda_-\over\sqrt{(\lambda_+-\lambda_-+2h)
(\lambda_+-\lambda_--2h)}}
\end{equation}
If we call this Jacobian $J(x,t)$, the partition function for this systems
reads
\begin{equation}
Z = \int\left[\prod_{x,\tau} d\lambda_\pm\,d{\bf m}_\pm\,J(x,\tau)\right]~
\exp-S_E
\end{equation}
In order to bring this expression into its usual form, we must bring the
Jacobian within the exponential:
\begin{equation}
\prod_{x,\tau}J(x,\tau) = \exp-W[\lambda_\pm]
\end{equation}
where
\begin{equation}
W[\lambda_\pm] = \int dx\,d\tau~{1\over 2a^2}\ln\left[{
(\lambda_+-\lambda_-)^2-4h^2\over
(\lambda_+-\lambda_-)^2}\right]
\end{equation}
Here we have introduced the lattice spacing $a$ in the imaginary time
direction as well as in the spatial direction, in order to incorporate
the functional Jacobian into a local action.\cite{note1}
We may now proceed to the functional integration of the fields ${\bf m}_\pm$,
which is done exactly like in the ordinary nonlinear $\sigma$ model, except
that $\sigma$ is replaced by $\lambda_\pm$. The effective potential in the
large-$N$ limit is then
\begin{eqnarray}
V[\lambda_\pm] &=& {\lambda_++\lambda_-\over2g} -
{1\over 2a^2}\ln\left[{
(\lambda_+-\lambda_-)^2-4h^2\over
(\lambda_+-\lambda_-)^2}\right] \nonumber\\
&& -\frac32\sum_{i=\pm}\int {dk\over2\pi}{d\omega\over2\pi}
\ln\left[ 1 + {\lambda_i\over k^2+\omega^2}\right]
\end{eqnarray}
We now want to solve the saddle-point equation for this potential.
To this end, let us write $\lambda_\pm = \alpha\pm\beta$. The saddle-point
equations are
\begin{mathletters}
\begin{eqnarray}
{\partial V\over\partial\alpha }&=& 0 = \frac1g +\frac3{4\pi}
\ln {\sqrt{\alpha^2-\beta^2}\over\Lambda^2 }\\
{\partial V\over\partial\beta }&=& 0 = -{h^2\over a^2\beta(\beta^2-h^2)}
 + \frac3{8\pi}\ln{\alpha+\beta\over\alpha-\beta}
\end{eqnarray}
\end{mathletters}
Here again $\Lambda$ is a momentum cutoff of the order of $1/a$.
In the decoupled limit ($h=0$) the solution to these equations is
$\beta=0$ and $\alpha=\alpha_0=2\Lambda\exp-4\pi/3g$. If we now define
reduced variables $\bar\alpha=\alpha/\alpha_0$, $\bar\beta=\beta/\alpha_0$
and $\bar h=h/\alpha_0$, we can put the above saddle-point equations
in the following form:
\label{gap3}
\begin{mathletters}
\begin{eqnarray}
&& \bar\alpha^2 - \bar\beta^2 = 1\\
&& \bar\beta(\bar\beta^2-\bar h^2)
\ln{\bar\alpha+\bar\beta\over\bar\alpha-\bar\beta} = \eta^2\bar h^2
\end{eqnarray}
\end{mathletters}
where $\eta^2 = 8\pi/3\alpha_0 a^2$. The gap at finite inter-chain
coupling is then $\Delta(K)=\Delta(0)\sqrt{\bar\alpha-\bar\beta}$.
The equations (\ref{gap3}) may be solved numerically. We need some
input about the parameter $\eta$: using the saddle-point value for
$\alpha_0$, one finds $\eta\sim 14.1$. A numerical solution of
Eq.~(\ref{gap3}) is illustrated on Fig.(\ref{fig2}), along with the
relation (\ref{gapM1}). Both of these results are obtained in the
saddle-point approximation, but in terms of different variables.
Of course, the change of variables defined in Eq.~(\ref{eigen}) should
not make any difference if the exact value of the gap could be calculated;
the $1/N$ expansions in terms of $\lambda_\pm$ and $\sigma_{1,2}$ are however
different.

\section{Numerical results}

In this section we present some numerical results obtained for the
spin-1 Heisenberg ladder ($s=\tilde s=1$) and compare them with the
analytical predictions worked out in the previous two sections of the
paper.

The system (\ref{Hamil1}) may be studied numerically by essentially two
techniques: exact diagonalization and quantum Monte-Carlo. Exact
diagonalizations may further be refined by applying the density-matrix
renormalization procedure. The disadvantage of exact diagonalizations is
of course that the amount of computer memory and computer time needed grows
exponentially with system size, whereas Monte-Carlo simulations are
time-consuming but may easily be implemented on medium-size systems.

Before presenting results, let us briefly describe the method. We used a
projector Monte-Carlo (PMC) technique, adapted from the work of
Takahashi\cite{Takahashi89}, with slight modifications. The method requires
that the off-diagonal elements of the Hamiltonian be non positive. The
Heisenberg Hamiltonian (\ref{Hamil1}) does not immediately fulfills this
condition, but on a bipartite lattice such as the ladder, a unitary
transformation $H\to UHU^{-1}$ may be defined that brings the Hamiltonian
into this form, with
\begin{equation}
\label{Utransfo}
U = \exp i\pi\sum_{i~\rm even}\left[ S_i^z + \tilde S_{i+1}^z \right]
\end{equation}
Let us still denote the resulting Hamiltonian by $H$.

The Hilbert space of the system may be described by the basis of
eigenstates of $\{S_i^z,\tilde S_i^z\}$: let us write such a basis
state as $|\{s_i\}\rangle$. The essence of the PMC method is the discrete
(imaginary) time evolution of an ensemble of {\it walkers}. A walker is
a state that belongs the the basis described above, changing from time to
time according to stochastic transitions. Each walker $|n\rangle$ is
assigned a non negative weight $w_n(\tau)$ which depends on time. The
walkers are evolved by repeated application of the operator
$W=\exp-\epsilon H$, where $\epsilon$ is some small time interval. This
evolution operator is applied stochastically, meaning that upon application,
a state $|\{s_i\}\rangle$ is changed to an other state $|\{s'_i\}\rangle$
with a probability proportional to the matrix element
$\langle\{s'_i\}|W|\{s_i\}\rangle$. At the same time, the weight $w_n$ is
multiplied by the diagonal element $\langle\{s_i\}|W|\{s_i\}\rangle$.
Of course, the exponentiation of $H$ is impossible to achieve exactly
without knowing the exact solution to the problem. We therefore divide
the Hamiltonian into different parts, each being the sum of mutually
commuting two-spin Hamiltonians. For the ladder, one needs three such parts,
as follows:
\begin{mathletters}
\begin{eqnarray}
H_1 &=& J\sum_{i~\rm even} \left[ {\bf S}_i\cdot{\bf S}_{i+1} +
{\bf\tilde S}_i\cdot{\bf\tilde S}_{i+1}\right] \\
H_2 &=& J\sum_{i~\rm odd} \left[ {\bf S}_i\cdot{\bf S}_{i+1} +
{\bf\tilde S}_i\cdot{\bf\tilde S}_{i+1}\right]\\
H_3 &=& K\sum_{i} {\bf S}_i\cdot{\bf\tilde S}_{i}
\end{eqnarray}
\end{mathletters}
The discrete evolution operator is then replaced by
$W=W_1W_2W_3$, where $W_i=\exp-\epsilon H_i$. Since the different terms
in $H_i$ commute, we only need to exponentiate a two-spin interaction, which
can be done analytically. The smaller $\epsilon$ is, the smaller the error
made in this approximation to the real exponential.

The iteration starts with a random distribution of $N$ walkers, all with equal
weights (e.g. $w_n=1$). More precisely, the walkers are chosen randomly
within the subspace of interest, i.e., for a fixed value of $S^z_{tot}$.
The number $N$ is typically 10~000. At each step the imaginary time is
increased by $\epsilon$ (typically $\epsilon\sim 0.05~J$) and the operator $W$
is applied as described above. After a few iterations the distribution of
weights becomes too sparse (its standard deviation becomes too large in
comparison with the mean) and the weights have to be reconfigured by a
procedure eliminating the ligth walkers (see Refs.~\onlinecite{Takahashi89}
and \onlinecite{Hetherington83}) and normalizing the weights. This
reconfiguration procedure is applied regularly, but should not affect the
physical properties of the ensemble when done properly. After the system has
evolved for some time  $\tau=n_\tau\epsilon$, the energy of the ensemble of
walkers is evaluated by projecting onto some arbitrary state $|\xi\rangle$.
For convenience, this state is taken to have an equal projection on each of
the basis states $|\{s_i\}\rangle$. The evolution then continues, an energy
measurement begin taken every $n_\tau$ iterations. The time interval
$n_\tau$ should be short enough to allow as many measurements as possible, but
long enough for successive energy measurements to be statistically
uncorrelated.
\begin{figure}[tpb]
\vglue 0.4cm\epsfxsize 8.7cm\centerline{\epsfbox{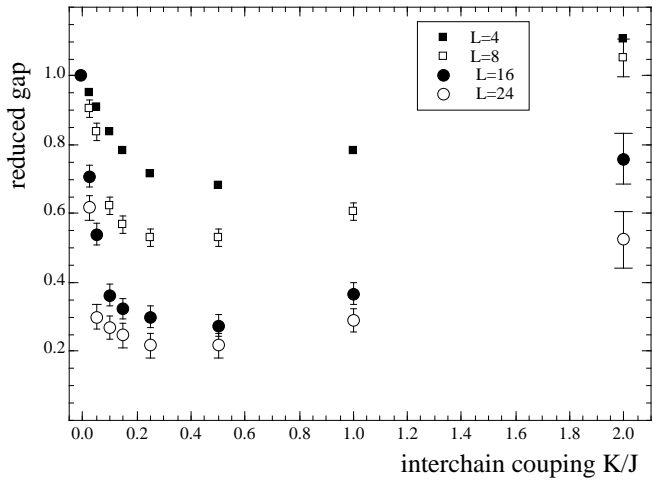}}\vglue 0.4cm
\caption{
Results of a Quantum Monte Carlo simulation for the gap $\Delta$
divided by the $K=0$ value $\Delta_0$.
The data are shown for $N=4$, 8, 16 and 24, where $N$ is the length of the
ladder. $\Delta_0$ was also obtained from exact diagonalizations, when
available.}
\label{fig3}
\end{figure}
In order to measure the spin excitation gap this way, one first finds the
lowest energy level in the $S^z_{tot}=0$ sector, and then in the
$S^z_{tot}=1$ sector. The difference should be equal to the gap $\Delta$.
Notice that the unitary transformation (\ref{Utransfo}) has effectively
shifted the momentum of excitations by $\pi$, so that the first excited
state of a Haldane system has now momentum zero and is not eliminated
by projecting onto the uniform state $|\xi\rangle$.

The outcome of the simulation for the gap and the ground state energy per
site as a function of $K$ is presented on Figs. 3 and 4. Notice that the
gap exhibits a sharp drop until about $K\sim 0.5$, after which it rises
slowly. We have not performed simulations on ladders longer than 24 rungs
(48 sites), which is not enough to do a satisfactory finite-size analysis
for most values of $K$. Nevertheless, we feel that the correct qualitative
behavior as a function of $K$ may be extracted.
\begin{figure}[tpb]
\vglue 0.4cm\epsfxsize 8.7cm\centerline{\epsfbox{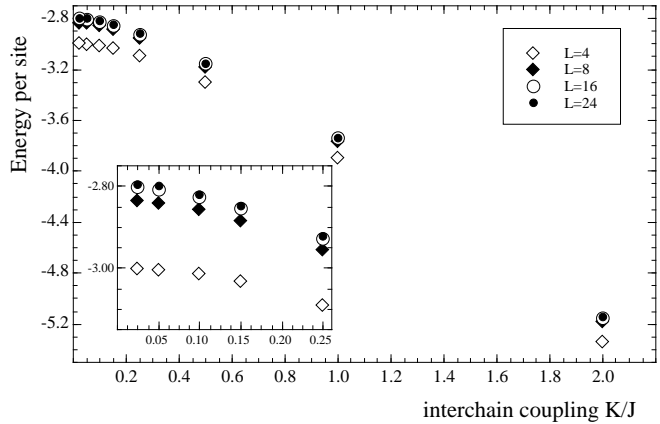}}\vglue 0.4cm
\caption{
Results of a Quantum Monte Carlo simulation for the ground state energy per
site $\varepsilon_0=E_0/N$, with $N=4$, 8, 16 and 24.}
\label{fig4}
\end{figure}

\section{Discussion}

In the first part of this work we mapped the Hamiltonian of the Heisenberg
ladder to the one-dimensional nonlinear sigma model, within the path integral
formalism and the approximation of local AF order.
{}From our knowledge of the nonlinear sigma model (with or without topological
term), we concluded that the ladder has an excitation gap for all values of
the inter-chain coupling $K$ if the spins $s$ and $\tilde s$ of the two chains
are both integers, or both half-integers (i.e. if $s+\tilde s$ is an integer),
whereas it has no gap if $s+\tilde s$ is a half-integer. In the former case,
the ladder is in a Haldane phase for all nonzero values of $K$, positive or
negative. This agrees with the interpretation of Ref.~\onlinecite{White95}

However, the quantitative estimate (\ref{gap2}) of the gap $\Delta$ as a
function of $K$, obtained via the saddle-point approximation of the nonlinear
sigma model and illustrated on Fig.~1, is clearly wrong in both the $K\to0$
and $K\to\infty$ limits. For small $K$, it announces a uniform increase of the
gap from a finite value at $K=0$. In reality, the gap vanishes as $K\to0$ if
$s=\tilde s=\frac12$, whereas it drops from its $K=0$ value if $s=\tilde
s=1$.\cite{note2} Thus, there is a qualitative difference between
half-integer and integer spin in the ladder when $K$ is small, as we
naturally expect. This difference does not show up in the nonlinear sigma
model representation of the ladder.

For large values of $K$, Eq.~(\ref{gap2}) predicts a square-root behavior
$\sim\sqrt{K}$ when $K$ is large. In reality, the gap is approximately equal
to $K$ in that limit, i.e., the energy difference between a singlet and a
triplet on a single rung, when $J$ is neglected before $K$. In fact, we do not
expect the parameters of the ladder non-linear sigma model of Sect.~II
to be correct when the ratio $K/J$ is too far from unity.
However, for moderate values of $K$ ($0.5 < K < 2$), the gap of the spin-1
ladder illustrated in Fig.~3 shows a moderate increase whose slope is
compatible with the estimate of Fig.~1.

The behavior of the ladder for small inter-chain coupling ($K\ll J$) is
better represented by two interacting nonlinear sigma models. In the
half-integer spin case, the presence of a topological term in each of the
chain makes the analysis quite difficult; we hope to report on this in later
work. However, the integer-spin case lends itself to a saddle-point
evaluation of the gap which is much closer to reality. In the absence of
variational principle or other guideline, it is difficult to say which
of the two saddle-point methods illustrated on Fig.~2 offers a better
estimate of the gap. A comparison with the Monte Carlo data of Fig.~3
tilts the balance in favor of the lower curve, which displays a sharper
drop as $K$ increases from zero. Recall that the saddle-point approximation
leading to this curve  neglects the fluctuations of the mass matrix $\Sigma$
of Eq.~(\ref{massMatrix}), whereas the upper curve is obtained by neglecting
the fluctuations of the Lagrange multipliers $\sigma_i$ of
Eq.~(\ref{coupledLag}).

\acknowledgements
The author thanks L.~Chen and A.-M.~Tremblay for discussion.
Financial support from the Natural Sciences and Engineering Research Council
of Canada (NSERC) and le Fonds pour la Formation de Chercheurs et l'Aide \`a
la Recherche du Gouvernement du Qu\'ebec (FCAR) is gratefully acknowledged.



\begin{references}
\bibitem[*]{note1} Choosing the same lattice spacing in the imaginary-time and
space directions is somewhat arbitrary, but does the least violence to the
Lorentz invariance of the theory.
\bibitem[**]{note2} If the numerical results of Ref.~\onlinecite{Barnes93}
for finite-length ladders were plotted in terms of $\Delta/\Delta_0$, as in
Fig.~3, we would observe a rise of the curve as the length $L$
increases, in contrast to Fig.~3.
\end{references}
\end{document}